# $Nb_2SiTe_4$: A Stable Narrow-Gap Two-Dimensional Material with Ambipolar Transport and Mid-Infrared Response

*Mingxing Zhao[1,3,4†], Wei Xia[1,4†], Yang Wang[2], Man Luo[2], Zhen Tian[1], Yanfeng Guo[1*], Weida Hu[2*] and Jiamin Xue[1*]*

[1]School of Physical Science and Technology, ShanghaiTech University, Shanghai 201210, China

[2]Shanghai Institute of Technical Physics, Shanghai 200083, China

[3]Shanghai Institute of Ceramics, Shanghai 200050, China

[4]University of Chinese Academy of Sciences, Beijing 100190, China

ABSTRACT

Two-dimensional (2D) materials with narrow band gaps (~ 0.3 eV) are of great importance for realizing ambipolar transistors and mid-infrared (MIR) detections. However, most of the 2D materials studied so far have band gaps that are too large. A few of them with suitable band gaps are not stable under ambient conditions. In this study, the layered $Nb_2SiTe_4$ is shown to be a stable 2D material with a band gap of 0.39 eV. Field-effect transistors based on few-layer Nb$_2$SiTe$_4$ show ambipolar transport with similar magnitude of electron and hole current and high charge-carrier mobility of ~ 100 cm$^2$V$^{-1}$s$^{-1}$ at room temperature. Optoelectronic measurements of the devices show clear response to MIR wavelength of 3.1 μm with a high responsivity of ~ 0.66 AW$^{-1}$. These results establish Nb$_2$SiTe$_4$ as a good candidate for ambipolar devices and MIR detection.

TEXT

Materials with narrow band gaps in the range of 0.15 eV to 0.4 eV hold special position in the family of semiconductors for their applications in mid-infrared (MIR) detections. Among them, indium arsenide, mercury cadmium telluride and so on are the most studied and used materials. Besides these conventional semiconductors, newly developed narrow-gap two-dimensional (2D) materials are also showing their great potential in the MIR detections. For example, few-layer black phosphorus (bP) with a band gap of ~ 0.3 eV has been used to demonstrate high photo responsivity and detectivity in the MIR range.[1-3] When alloyed with arsenide, the band gap of black arsenic phosphorus (bAP) can be further tuned from ~ 0.3 eV to ~ 0.15 eV, which extends the spectrum to the long-wavelength infrared.[4-5] The layered nature of these materials gives them large flexibility and ease of integration with other materials to form various device structures.[3, 6-7]

Besides their applications in optoelectronics, narrow-gap 2D semiconductors such as bP can be used as the channel material of field-effect transistors (FET) and support ambipolar transport, in which the majority carriers of the channel can be easily switched between electrons and holes by electrostatic doping of the gate voltage.[8-9] This property offers great opportunity for novel device functions such as field-programmable p-n junctions,[10] which are not achievable by conventional semiconductors.

Due to these unique applications, narrow-gap 2D materials are highly demanded. However, most of the 2D semiconductors studied so far have band gaps in the order of 1 eV or larger.[11] bP and bAP have suitable band gaps, but degrade within hours under ambient condition due to the chemical instability of phosphorus.[4, 12-13] Thus new stable narrow-gap 2D materials are needed. Here we report a layered material $Nb_2SiTe_4$ (NST) with a band gap of 0.39 eV. When few-layer NST is fabricated into an FET, it shows ambipolar transport with similar capability of conducting electrons and holes. A respectable hole mobility of ~ 100 $cm^2V^{-1}s^{-1}$ at room temperature is obtained. When illuminated by MIR light with 3.1 μm wavelength, NST devices demonstrate high photo responsivity of ~ 0.66 $AW^{-1}$. They also show good stability under ambient condition without any encapsulation. These findings demonstrate NST to be a stable narrow-gap 2D material with great application potentials.

RESULTS AND DISCSSION

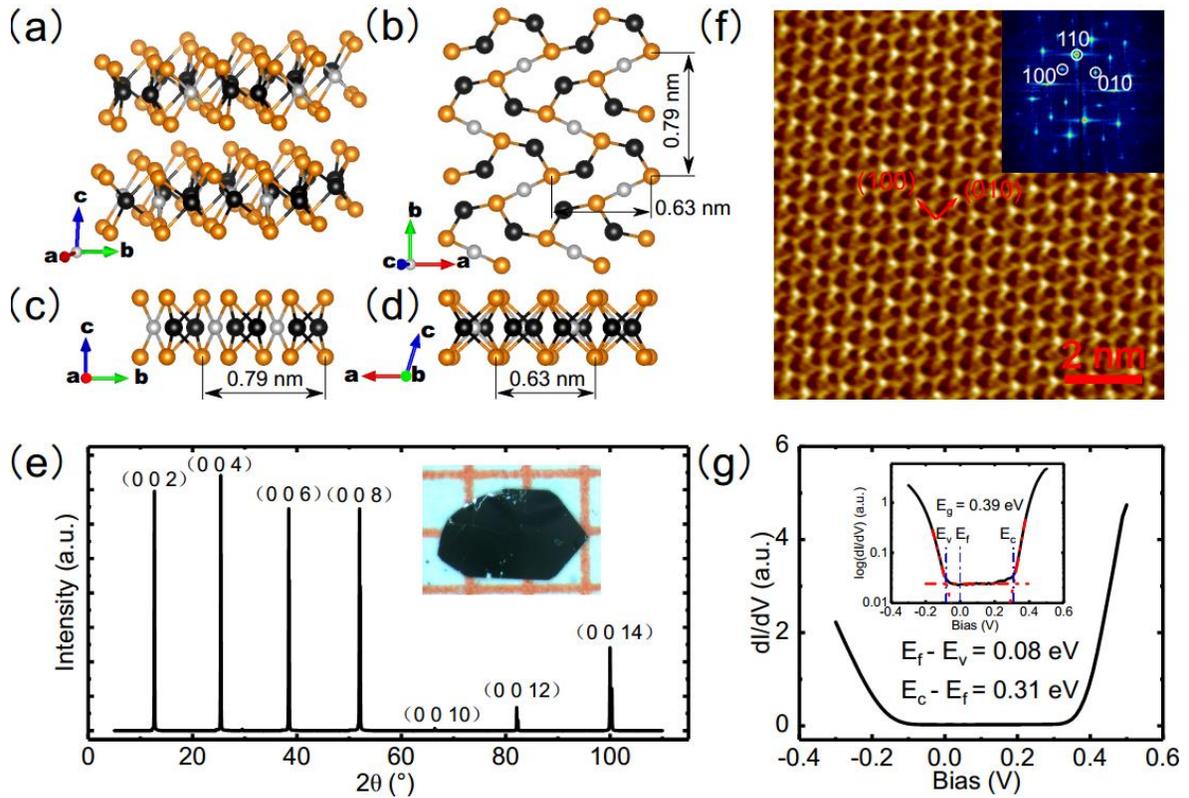

Figure 1. (a) Perspective view of few-layer NST showing atoms of Te (gold), Nb (black) and Si (gray). (b) Top view and (c), (d) side views of monolayer NST. (e) XRD data of NST. The X-ray source is Cu Kα line with the wave length of 0.154 nm. Inset: optical image of a bulk NST crystal. (f) Atomically resolved image of a NST crystal measured by STM with the corresponding fast Fourier transform in the inset. STM parameters: $V = 0.5$ V, $I = 100$ pA. (g) The d$I$/d$V$ spectra of a NST crystal. Inset is the same data plotted in log scale for a better determination of band edges. The band gap $E_g$ is measured to be 0.39 eV. STS parameters: $V_{initial} = 0.5$ V, $I_{initial} = 100$ pA; lock-in frequency $f = 991.2$ Hz, modulation voltage $V_{rms} = 10$ mV.

NST has been synthesized decades ago,[14] but there have been scarce studies about it. Bulk NST crystal belongs to the space group of *P121/c1* (No. 14),[15] which is in the monoclinic crystal system. Figure 1a to 1d show the different views of the NST structure. It is a layered material formed by stacking the Te-(Nb, Si)-Te sandwich layers, which bears similarity to the transitional metal dichalcogenides (TMDs) such as $MoS_2$,[16] $NbSe_2$[17] and so on. The inner atomic layer consists of two types of atoms (Nb and Si) instead of one in the TMDs. The lattice parameters have been determined to a = 0.63 nm, b = 0.79 nm, c = 1.47 nm, α = 90°, β = 107°, γ = 90°.[15] We synthesized NST crystal with self-flux method (see the Experimental section). The as grown crystals are black thin flakes with metal luster and sizes of millimeters (inset of Figure 1e). Due to the layered nature, powder X-ray diffraction (XRD) mainly probes the basal planes in the crystal and the layer distance is extracted to be 0.697 nm which gives a lattice constant c = 1.46 nm. Sharp Bragg peaks indicate the high quality of our crystals. To complement XRD study, we used scanning tunneling microscope (STM) which resolves the atomic structure within the basal plane (Figure 1f). The orthorhombic structure is clearly visible with lattice constants of 0.63 nm and 0.78 nm, agreeing very well with literature values.[15] From Figure 1 b and f we can see that a unit cell of NST contains a large number of atoms, which gives many phonon modes as shown by Raman spectroscopy (Figure S1 in the Supporting Information, SI). Besides revealing the atomic structure, more importantly, STM can be used to measure the band structure with the technique of scanning tunneling spectroscopy (STS).[18-19] In Figure 1g, The Fermi level is located at 0.08 eV above the valence band edge $E_v$, indicating its intrinsic p-type doping. The band gap $E_g$ is determined to be 0.39 eV, in the MIR range, which is promising for both ambipolar transport and MIR detection.

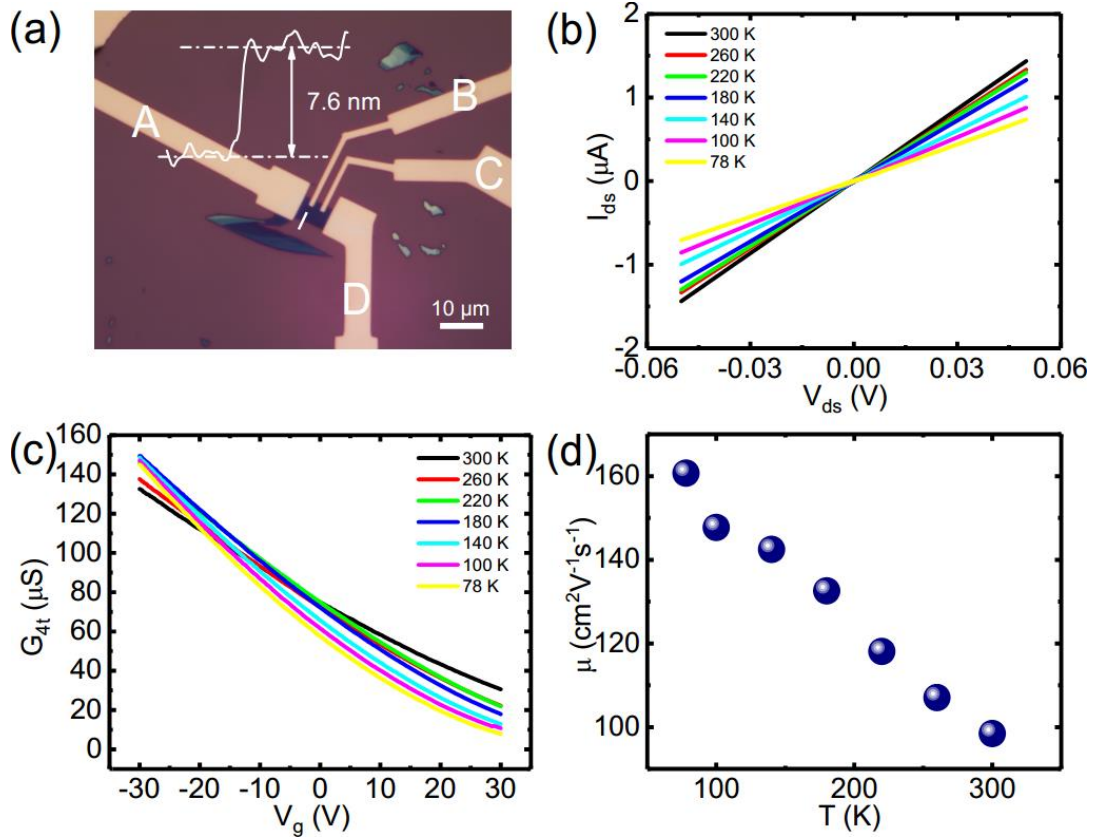

**Figure 2.** (a) The optical image of the few-layer NST device with the height profile of the channel along the white line. (b) Drain-source current $I_{ds}$ between electrodes A and D versus $V_{ds}$ at different temperatures. (c) Four-probe conductance ($G_{4t}$) as a function of back-gate voltage ($V_g$) at various temperatures. (d) Four-probe hole mobility as a function of temperature extracted from (c).

To measure the transport property of few-layer NST, scotch tape method is adopted to exfoliate NST onto highly p-doped Si substrate covered with 300 nm thermally grown $SiO_2$. Then electron-beam lithography and metal deposition are used to fabricate the device. A typical FET device is shown in the optical image of Figure 2a. The flake thickness is measured by atomic force microscope (AFM) to be 7.6 nm (other thicknesses of few-layer NST showed similar behavior). The metal electrodes consist of 10 nm of Ti and 40 nm of Au. All transport measurements were performed in vacuum (~$10^{-5}$ mbar) and in dark with a Janis ST-500 probe station.

The linear two-terminal output curves measured from 300 K to 78 K (Figure 2b) indicate that Ti forms good contacts with NST. We determine the work function of NST to be 4.36 eV by ultraviolet photoelectron spectroscopy (Figure S2), which matches well with that of Ti (4.3 eV).[20] To extract the intrinsic field-effect mobility, four-terminal transfer curves are measured as a function of temperature (Figure 2c). The p-type transport is consistent with the band structure measured by STS in Figure 1g. Mobility $\mu$ is calculated by $= \frac{L}{WC_{ox}}\frac{dG}{dV_g}$, where $G$ is the conductance, $V_g$ is the back-gate voltage, $C_{ox} = 1.15 \times 10^{-8}$ F cm$^{-2}$ is the specific capacitance of gate oxide, and $L$ ($W$) is the length (width) of the channel. The mobility at room temperature is 98 cm$^2$V$^{-1}$s$^{-1}$ and it increases as temperature is lowered, reaching 160 cm$^2$V$^{-1}$s$^{-1}$ at 78 K (Figure 2d). Compared with the majority of 2D materials, NST has a higher mobility,[21-22] which is essential for high performance electronic devices. Similar transport properties have been obtained on 8 devices (see Figure S3 for data of two other devices), although some variations in the transport properties such as mobility and threshold voltage exist among them. With device optimization, such as better substrates and dielectric screening,[23-24] the mobility should be further improved.

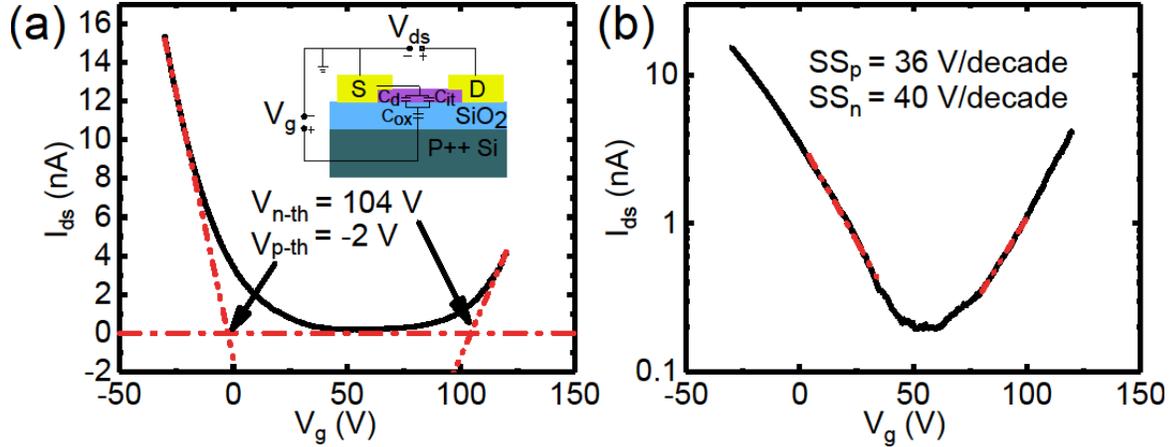

**Figure 3.** (a) Transfer curve of a NST device showing ambipolar transport. $V_{n\text{-th}}$ and $V_{p\text{-th}}$ represent the threshold voltages on the electron side and hole side, respectively. Inset: schematic of equivalent capacitance model of the device. $C_d$ and $C_{it}$ represent the depletion layer capacitance and interface capacitance, respectively. (b) Data from (a) plotted in log scale. Red lines are the linear fitting curves to extract the subthreshold swings of the electron side ($SS_n$) and hole side ($SS_p$).

Due to the narrow band gap, it is possible to change the majority carrier type in the channel from holes to electrons simply by the back-gate voltage. Figure 3a shows the transfer curve of a 5.7 nm NST FET (Figure S4), where ambipolar behavior can be clearly seen. The electron and hole currents have comparable magnitude, indicating the similar Schottky barrier heights between Ti and the conduction and valence bands of NST. Previously reported narrow-gap 2D materials such as bP can also support ambipolar transport,[8-9] but the electron current is usually orders of magnitude lower than the hole current, which undermines the tunability of a device.

Further analysis of the data in Figure 3a can be used to extract the transport gap, which is given by[25]

$$E_g = e\left(V_{ds} + \left(\frac{V_{n-th}}{\beta_n} - \frac{V_{p-th}}{\beta_p}\right)\right), \quad (1)$$

where $e$ is the electron charge, $V_{ds}$ is the source-drain bias (100 mV in Figure 3a), $V_{n-th}$ ($V_{p-th}$) is the threshold voltage on the electron (hole) side, and $\beta_n$ ($\beta_p$) is the gate efficiency in tuning the band edge on the electron (hole) side, which is defined below. The threshold voltages are determined from Figure 3a. The subthreshold swing (SS) of a FET can be written as[26]

$$SS = \ln 10 \frac{kT}{e}\left(1 + \frac{C_d + C_{it}}{C_{ox}}\right) \equiv \ln 10 \frac{kT}{e}\beta, \quad (2)$$

where $k$ is the Boltzmann constant, $T$ is the temperature (150 K in our case), and $C_d$ ($C_{it}$) is the depletion (interface) capacitance. The gate efficiency $\beta$ is defined as $\beta = 1 + \frac{C_d + C_{it}}{C_{ox}}$. The equivalent capacitance circuit is shown in the inset of Figure 3a. Due to the presence of $C_d$ and $C_{it}$, the gate voltage $V_g$ in the subthreshold regime can only cause a down scaled shift of the bands in the device. The scaling factor $\beta$ can be extracted from the fittings in Figure 3b. Since the electron side and hole side show slightly different SS, both values are used in Equation 1. The calculated transport gap of NST is 0.18 eV, less than the STS value. This difference may come from the thermal broadening of the band edges and the non-uniform doping in the device.

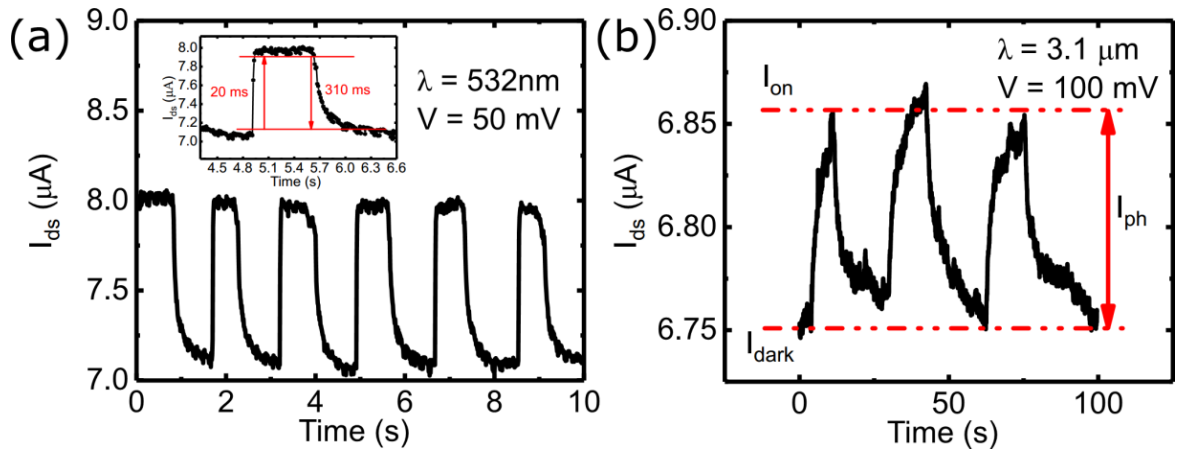

**Figure 4.** (a) Response to laser pulses with 532 nm wavelength. The rising and falling time is defined in the main text. (b) Response to MIR laser pulses with 3.1 μm wave length.

The band gap values of NST measured by STS and transport indicate that NST devices should have MIR responses. To test its optoelectronic property, we first use 532 nm laser pulses to characterize the NST FETs as shown in Figure 4a. With 50 mV source-drain bias, 0 V gate voltage and 310 μW laser, the device shows ~ 1 μA of photocurrent and fast response. We define the rising and falling time as the time span between 90% and 10% of the total photocurrent, which gives 20 ms rising edge and 310 ms decaying edge. We note that the dark current is high (~ 7 μA) due to the intrinsic heavy p type doping. Response to MIR wavelength (3.1 μm) is also measured as shown in Figure 4b, with the laser power of ~160 nW. The photocurrent is about 100 nA which gives a responsivity of 0.66 A W$^{-1}$, much larger than the responsivity of bAP.[4] These results demonstrate the potential of NST for MIR detection.

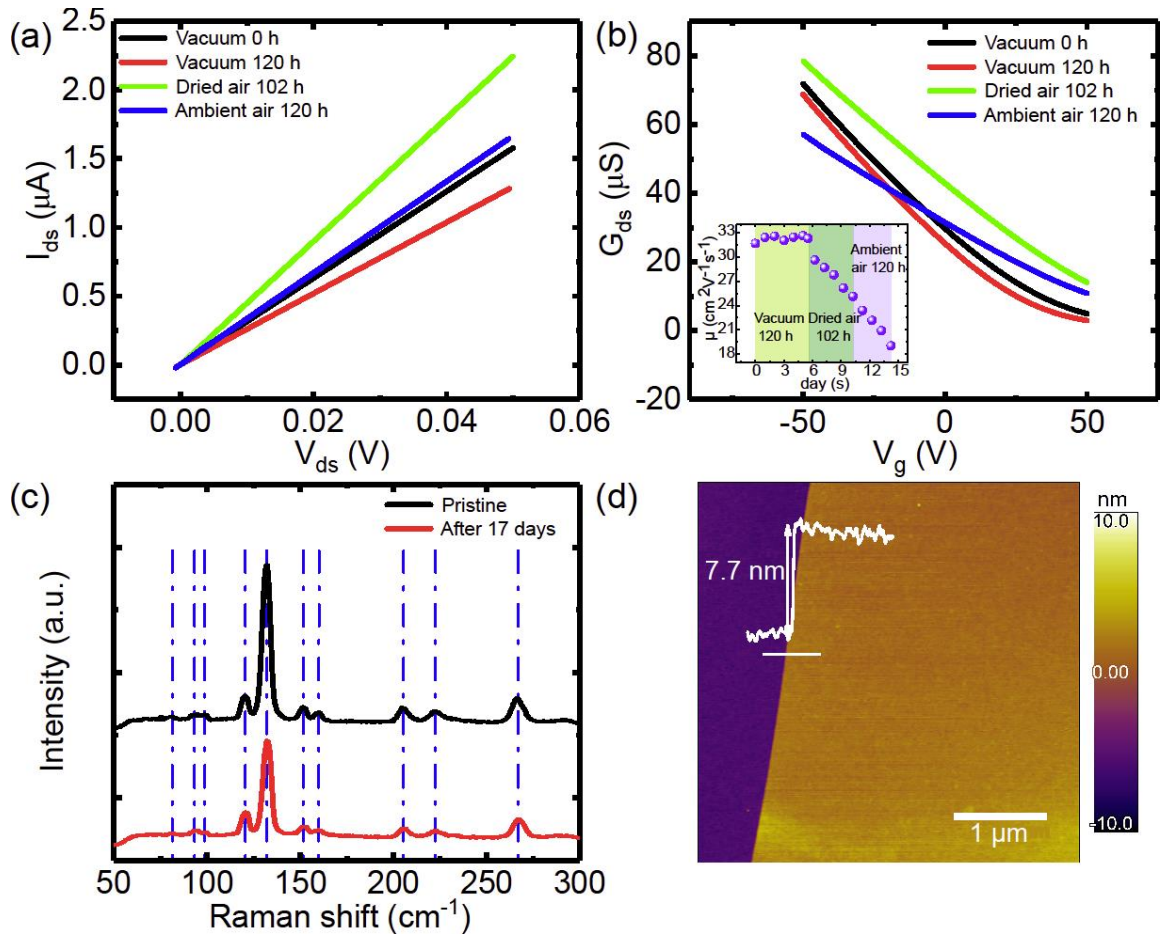

**Figure 5.** (a) $I_{ds}$ versus $V_{ds}$ under different test conditions. (b) Transfer curves show the slight degradation of mobility after the device is exposed to ambient air for 120 hours. Inset: the day-to-day variation of the device mobility. (c) Raman spectra taken before and after the transport measurements shown in (a) and (b). (d) AFM image of the channel after the transport measurements showing the smooth intact surface.

The ambipolar transport and MIR response have shown the application potential of NST. Stability is another important property we should examine as it has always been a great challenge for narrow bandgap semiconductors, such as bP and bAP.[4, 12] These materials tend to react quickly with oxygen and water moisture in air and degrade completely in the time scale of minutes to several hours.[27] Exfoliations of these materials are usually done in glove boxes with water and oxygen contents less than 1 ppm. Special care during device fabrication and various encapsulation techniques are needed to protect them,[4, 13, 23, 28] which bring in extra difficulties and complexities. However, NST shows remarkable resistance to ambient conditions. Exfoliation and inspection are all carried out in air. As shown in the transport data in Figure 5a and 5b, when a fabricated device is stored in vacuum for 120 hours, the mobility increases slightly (from black to red curve), due to the desorption of surface contaminant. When exposed to dried air for another 102 hours, mobility is slightly decreased with an increase of conductance (green curve), presumably due to the mild oxidation effect from the air. When the device is stored in air with 40% relative humidity for another 120 hours, the device still shows good transport behavior (blue curve), with 40% reduction of its mobility. In the inset of Figure 5b, day-to-day variation of the mobility is monitored. After exposed to air, the mobility decreases linearly without obvious change of speed when the air is switched from dry to ambient air with moisture, which indicates the key role of oxygen instead of water. As a comparison, bP and bAP devices usually completely lost their electrical conductivity after a few hours exposure to moisture air.[28] Raman spectra of the same device before and after measurements also reveal the stability of NST (Figure 5 c). All the peaks remain their positions and line profiles, and no new peaks can be seen. The decrease in intensity can be ascribed to the surface oxidization of NST. The AFM image (Figure 5d) of the channel after all the characterizations shows the cleanness and flatness of the flake, as compared with the chemically-corroded rough surface of bP and bAP.[4, 12, 29] A similar stability can be seen on another device in Figure S5 of the SI.

CONCLUSIONS

In conclusion, we have explored the layered material $Nb_2SiTe_4$ as a 2D narrow-gap semiconductor. Ambipolar transport with respectable mobility and comparable electron-hole injection indicate its outstanding electrical property. Good responsivity to MIR wavelength demonstrates the great potential for optoelectronic applications. Besides these, its superior stability renders $Nb_2SiTe_4$ a unique position among all the narrow-gap 2D materials, which warrants its further exploration.

METHODS

**Crystal growth :** $Nb_2SiTe_4$ crystals were grown by using Si:Te=1:4 as the flux. A mixture of Nb (99.95%, Aladdin), Si (99.999%, Aladdin) and Te (99.99%, Aladdin) in a molar ratio of 1:2:8 was placed into an alumina crucible. The crucible was sealed into a quartz tube in vacuum and then heated in a furnace up to 1150 °C in 15 hours. After reaction at this temperature for 5 hours, the assembly was slowly cooled down to 750 °C and stayed at 750 °C for more than 10 hours. The excess Si and Te was quickly removed at this temperature in a centrifuge. Thin pieces of black $Nb_2SiTe_4$ flakes with metallic luster were obtained at the alumina crucible.

**STM and STS:** The STM experiments were carried out in the Omicron LT-Qpuls-STM with tungsten tip calibrated on Ag(111) surface. To get clean surface, the NST bulk sample was exfoliated in ultrahigh vacuum chamber with the base pressure $< 10^{-9}$ mbar and then transferred to STM chamber to start the scanning with the base pressure of $10^{-11}$ mbar and temperature of 77 K.

**Device fabrication and transport measurements:** All the FET devices in this study start with NST exfoliation on highly doped silicon wafer covered with 300 nm $SiO_2$. The electron beam lithography is used to design the contacts. Thermal evaporation of electrode metals (10 nm Ti and 40 nm Au) with lift off method is used to deposit the contacts. Then the devices are ready for transport measurement, which is carried out in a Janis ST-500 probe station under vacuum. Keithley source meters 2612B provide the source-drain voltage and back gate voltage while measure the current.

**XRD:** Powder X-ray diffraction was performed on $Nb_2SiTe_4$ single crystals aligned along the (001) plane, using a D8 Venture Bruker at room temperature. A Cu kα source (1.5418 Å) was used.

**Raman:** A custom made micro Raman system with ANDOR SR-500i-D2-R spectrometer and 1800 grooves per mm grating was used. The incident laser wavelength was 532 nm.

**AFM:** All AFM images were obtained by AFM (Oxford, Cypher S) under ambient conditions.

ASSOCIATED CONTENT

**Supporting Information**.

The following files are available free of charge. Additional figures and data to support the results in the main text (PDF).

AUTHOR INFORMATION

**Corresponding Authors**

*xuejm@shanghaitech.edu.cn, wdhu@mail.sitp.ac.cn, guoyf@shanghaitech.edu.cn

**Author Contributions**

The manuscript was written through contributions of all authors. All authors have given approval to the final version of the manuscript. †These authors contributed equally.

ACKNOWLEDGMENT

We thank Prof. B. Chen for initial help with the MIR measurements. The device fabrication was supported by the Centre for High-resolution Electron Microscopy (CℏEM), SPST, ShanghaiTech University under contract No. EM02161943. Drs. N. Yu, X. Wang and Z. Zou helped with the XRD. M. Z., Z.T. and J.X. are supported by the Thousand Talents Project and ShanghaiTech University. W. X. and Y.G. are supported by the Natural Science Foundation of Shanghai (17ZR1443300) and the Shanghai Pujiang Program (17PJ1406200). Y. W., M. L. and W. H. are supported by the National Natural Science Foundation of China (61674157, 11734016).